\documentclass[aps,prl,twocolumn,showpacs,groupedaddress]{revtex4}
\usepackage{graphicx}
\usepackage{dcolumn}
\usepackage{bm}

\draft

\begin{document}

\title{A possible route to grow a (Mn:Si$_{(1-x)}$Ge$_x$)-based diluted magnetic semiconductor}

\author{Ant\^onio J. R. da Silva}
\affiliation{Instituto de F\'\i sica, Universidade de S\~ao Paulo, CP 
66318, 05315-970, S\~ao Paulo SP, Brazil}
\author{Alex Antonelli}
\affiliation{Instituto de F\'{\i}sica, Unicamp, C.P. 6165,
13083-970 Campinas SP, Brazil}
\author{A. Fazzio}
\affiliation{Instituto de F\'\i sica, Universidade de S\~ao Paulo, CP 
66318, 05315-970, S\~ao Paulo SP, Brazil}

\begin{abstract}
We systematically investigate, using {\it ab initio} density-functional
theory calculations, the properties of interstitial and substitutional Mn in both Si and Ge, as well
as in the Si$_{1-x}$Ge$_x$ alloy. We show that volume effects are not the main reason Mn prefers to be a subsitutional
impurity in pure Ge, and chemical effects, therefore, play an important role.
Using realistic models of Si$_{1-x}$Ge$_x$, we show that for $x \gtrsim 0.16$ substitutional Mn in Ge-rich neighborhoods
become more stable than interstitial Mn, which may allow the growth of Si-based diluted magnetic semiconductors.
\end{abstract}

\date{\today}


\maketitle

Diluted magnetic semiconductor (DMS) materials \cite{dietl,science-dietl,ohno1,ohno2}
have been know for a long time \cite{book}, however, it was
the relatively recent growth of III-V \cite{ohno1,III-V} Mn-based ferromagnetic DMS that have brought these materials
to the spotlight, motivated by the possibility of fabricating useful spintronic devices.
Although these accomplishments are rather significant, if one considers the possible technological impact
there is no doubt that the synthesis of a Si-based DMS material would be of great relevance.
Ferromagnetism in the Mn$_x$Ge$_{1-x}$ compound has been reported by more than one group \cite{mnge},
which makes one ask the obvious question: would it be possible to grow a similar Mn$_x$Si$_{1-x}$
ferromagnetic material?

Considering this question in a recent paper \cite{nosso}, we have argued that:
(i) as Si and Ge have similar electronic structures, it is reasonable to expect that for large enough $x$, Mn$_x$Si$_{1-x}$
would become ferromagnetic below some critical temperature, in the same way Mn$_x$Ge$_{1-x}$ does it. This hypothesis is
supported by a recent calculation \cite{freeman} where both Mn$_x$Ge$_{1-x}$ and Mn$_x$Si$_{1-x}$ were studied.
The question then is if the required values of $x$ can be achieved; (ii) in Si, a Mn impurity
favors an interstitial site, whereas in Ge it prefers a substitutional site (see below); (iii) as a consequence, a Mn substitutional impurity in Ge cannot diffuse as easily as an
interstitial Mn in Si \cite{woodbury,difmnsi,fastdiff}, allowing the introduction of a large enough number of impurities
without their diffusion and subsequent clustering. Therefore, it seems that a crucial issue to obtain
a Mn$_x$Si$_{1-x}$ material is to find a way to have the Mn occupying substitutional sites. We have previously
found \cite{nosso} that there are substitutional sites at the Si(100) surface with formation energies comparable to
interstitial sites, opening up the possibility of placing Mn$_{Si}$ through some non-equlibrium growth
process, akin to the low-temperature molecular beam epitaxy procedures that are crucial to obtain III-V based DMSs \cite{III-V}.

In the present paper we address the following questions: (i) Is it simply a volume effect that would make Mn prefer a tetrahedral interstitial
site in Si whereas it prefers a substitutional site in Ge? (ii) Will a Ge rich environment in the SiGe alloy also favor a substitutional Mn
site? Regarding question (i), we find that the cause Mn is substitutional in Ge is not simply a volume effect, and therefore finding
ways to increase the Si lattice parameter would not suffice to stabilize the substitutional Mn impurity. However, since
specific chemical interactions are the possible explanation for the differences between Mn in Si and Ge, this implies that the answer
to question (ii) above is yes. We actually find that the substitutional Mn site shoud be the predominant one for Ge concentrations
in the Si$_{1-x}$Ge$_x$ alloy for $x \gtrsim 0.16$. These findings have the important consequence that a Mn doped SiGe alloy may
present similar ferromagnetic properties as the Mn$_x$Ge$_{1-x}$ material, opening up the possibility of Si-based spintronics.

All our results are based on {\it ab initio} calculations, based on spin-polarized density functional theory within the generalized gradient approximation 
(GGA) \cite{gga}. We have used ultrasoft pseudopotentials \cite{vanderbilt} and a plane wave expansion up to 230 eV, as implemented
in the VASP code \cite{vasp}. We have used a FCC-based supercell containing 128 sites, and the $L$-points for the Brillouin
zone sampling. In all calculations the positions of all atoms in the supercell were relaxed until all the forces components were smaller
than 0.02 eV/\AA.

We initially considered pure Si and Ge crystals. We placed a Mn atom both at an interstitial as well as
at a substitutional site. The formation energy of a neutral interstitial impurity \cite{neutral}, $E^I_f$, is calculated as 
\begin{equation}
E^I_f=E_{def}-E_{bulk}-\mu_{Mn}
\label{eform}
\end{equation}
where $E_{def}$ is the total energy of the supercell with the defect, $E_{bulk}$ is the total energy of
the supercell without the defect (either pure Si or pure Ge), and $\mu_{Mn}$ is the Mn chemical potential \cite{chempot}.
For neutral substitutional impurities \cite{neutral} the formation energy ($E^S_f$) is given by
\begin{equation}
E^S_f=(E_{def}+\mu_{X})-E_{bulk}-\mu_{Mn}
\label{eform2}
\end{equation}
where $\mu_{X}$ is the chemical potential of either Si or Ge \cite{chempot}.

In Table \ref{tab1}  we present the results for Mn atoms in Si and Ge at their equilibrium lattice constants
(we find $a_{Si}=5.445$ \AA\ for Si and $a_{Ge}=5.750$ \AA\ for Ge). As expected \cite{difmnsi,fastdiff}, for silicon the Mn atoms prefer
to be at interstitial sites, with an energy difference $\Delta E^{S-I}$ between the substitutional and interstitial sites
of $\Delta E^{S-I}=0.3$ eV. For Mn in Ge, on the other hand, we obtain that the Mn substitutional impurity has a lower
formation energy by $\Delta E^{S-I}=-0.6$ eV. In order to investigate how much of this difference between Si and Ge is a result
of their lattice parameter difference, {\it i.e.}, how much is due to a volume effect, we repeated the same calculations
for Si (Ge) in the lattice parameter of Ge (Si) \cite{chempot}. The results are also presented in Table \ref{tab1}.
In silicon, no significant changes were observed, with a small reduction in the substitutional Mn formation energy,
which led to $\Delta E^{S-I}=0.2$ eV instead of 0.3 eV. For Mn in Ge, the interstitial site becomes highly unfavorable,
with the formation energy changing by more than 1 eV. The formation energy for the substitutional site, on the other hand,
changed very little. This leads to a large increase in the energy difference between the substitutional and interstitial sites:
$\Delta E^{S-I}=-1.7$ eV. Therefore, even though increasing the Si lattice parameter has a small tendency towards favoring
a Mn atom in a substitutional site, {\it the difference in lattice parameter between Si and Ge cannot account, by itself, for the distinct behavior
of Mn in these materials}. Thus, specific chemical interactions between the Mn atom and the host atoms are crucial in determining
the distinct behavior in Si and Ge.
\begin{table}
\caption{Formation energies (in eV) for interstitial and substitutional Mn in
bulk Si and Ge. For both materials, the results are reported for the Si ($a_{Si}$) as well as the
Ge ($a_{Ge}$) lattice constants.}
\label{tab1}
\begin{tabular}{cccccc}
\hline
\hline
Lattice Constant & Mn$_I^{Si}$ & Mn$_{Si}$ & Mn$_I^{Ge}$ & Mn$_{Ge}$ \\
\hline
$a_{Si}$             &  2.5           & 2.8          &       3.4    & 1.7   \\
$a_{Ge}$            &  2.5           & 2.7          &       2.1    & 1.5   \\
\hline
\hline
\end{tabular}
\end{table}
This suggests that in the Si$_{1-x}$Ge$_x$ alloy the Mn atoms may prefer to be at a substitutional
site with a Ge rich environment, instead of at an interstitial site.
In order to confirm this possibility, we performed calculations for both interstitial as well as substitutional
Mn in Si$_{1-x}$Ge$_{x}$, for x=0.25, 0.5 and 0.75. Since the Si$_{1-x}$Ge$_{x}$ is a substitutionally disordered alloy,
the vicinity of a Mn impurity is not uniquely determined, and for either the substitutional or the interstitial
Mn, there are five different types of sites, if only the nearest-neighborhood
is considered, {\it i.e.}, a Mn surrounded by a configuration of Si and Ge atoms
that can be labeled as Si$^{\nu}$Ge$^{4-\nu}$, for $\nu$ varying from 0 to 4. 

To simulate the Si$_{1-x}$Ge$_{x}$ disordered alloys, we used supercells with 128 atoms where the
atoms were distributed as Special Quasirandom Structures (SQS) \cite{Wei}.
Details of the preparation of the supercells were given elsewhere \cite{liga1}, and
a similar procedure has already been used to study vacancies in this alloy \cite{liga2,liga3}.
It should be mentioned that due to the SQS approach, the disorder of the alloy is taken into
account in an explicit way. As we have shown before \cite{liga1}, the alloy lattice parameter
follows very closely the Vegard's law, and we therefore use $a_{Si_{1-x}Ge_{x}}=(1-x) a_{Si}+x a_{Ge}$.

To study the interstitial Mn defects, we randomly selected five sites with distinct first neighborhoods.
As these sites will also have different second, third, etc., neighborhoods, an averaging
procedure should be performed. From previous studies of vacancies in the alloy \cite{liga2,liga3},
we estimate an overall variation in the formation energies of the order of $\pm$ 0.1 eV due the different
vicinities, and hence this averaging will not alter our conclusions in any significative way.
The formation energy $E^I_{f}(\nu,x)$ of a neutral interstitial \cite{neutral} Mn in the Si$_{1-x}$Ge$_{x}$ alloy, in a given neighborhood 
Si$^{\nu}$Ge$^{4-\nu}$, labeled by $\nu$, is given by
\begin{equation}
E^I_{f}(\nu,x)=E_I(\nu,x)-E_{bulk}(x)-\mu_{Mn}(x).
\end{equation}
\noindent Here $E_I(\nu,x)$ is the total energy of the SQS structure with the Mn interstitital,
$E_{bulk}(x)$ is the total energy of the SQS alloy without any
defects, and $\mu_{Mn}(x)$ is the Mn chemical potential in the Si$_{1-x}$Ge$_{x}$ alloy, which
we determine \cite{chempot2} as
\begin{equation}
\mu_{Mn}(x)=(1-x) \mu_{MnSi}+x \mu_{MnGe}
-\mu_{Si_{(1-x)}Ge_{x}}.
\label{potmn}
\end{equation}

To study substitutional Mn defects a similar procedure as described above was employed.
Since an atom from the original SQS structure must now be removed, there is the extra possibility
of having the Mn replacing either a Si or a Ge atom. The formation energy $E^S_{f}(X,\nu,x)$ of a neutral \cite{neutral}
Mn substituting an atom $X$ ($X$=Si,Ge), in the Si$_{1-x}$Ge$_{x}$ alloy, with a neighborhood 
Si$^{\nu}$Ge$^{4-\nu}$, labeled by $\nu$, is given by
\begin{equation}
E^S_{f}(X,\nu,x)=E_S(X,\nu,x)-E_{bulk}(x)+\mu_{X}(x)-\mu_{Mn}(x).
\end{equation}
\noindent Here $E_S(X,\nu,x)$ is the total energy of the SQS structure with the Mn in place of an atom $X$ ($X$=Si,Ge),
which has a chemical potential $\mu_{X}(x)$ in the Si$_{1-x}$Ge$_{x}$ alloy \cite{chempot3}.
\begin{table}
\caption{Formation energies (in eV) for the interstitial and substitutional Mn in
Si$_{(1-x)}$Ge$_x$, for $x=0.25, 0.5$ and 0.75, and for the distinct neighborhoods
Si$^{\nu}$Ge$^{4-\nu}$. For the substitutional cases, Mn was always replacing
a Si atom, except for $x=0.5$ where Mn replacing Ge atoms were also considered (results in parenthesis).}
\label{tab2}
\begin{tabular}{c|cc|cc|cc}
\hline
\hline
             & \multicolumn{2}{c|}{x=0.25} & \multicolumn{2}{c|}{x=0.5} & \multicolumn{2}{c}{x=0.75} \\
$\nu$    & $E^I_{f}$ & $E^S_{f}$ & $E^I_{f}$ & $E^S_{f}$ & $E^I_{f}$ & $E^S_{f}$ \\
\hline
4       & 2.5 & 2.6 & 2.4 & 2.3 (2.2) & 2.2 & 2.0 \\
3       & 2.4 & 2.5 & 2.5 & 2.2 (2.2) & 2.2 & 1.9 \\
2       & 2.6 & 2.4 & 2.6 & 2.2 (2.1) & 2.5 & 1.8 \\
1       & 2.8 & 2.3 & 2.7 & 2.0 (2.1) & 2.5 & 1.8 \\
0       & 2.9 & 2.3 & 2.8 & 2.0 (2.0) & 2.6 & 1.7 \\
\hline
\hline
\end{tabular}
\end{table}

\begin{figure}[ht]
\includegraphics[angle=0,width=8.0 cm]{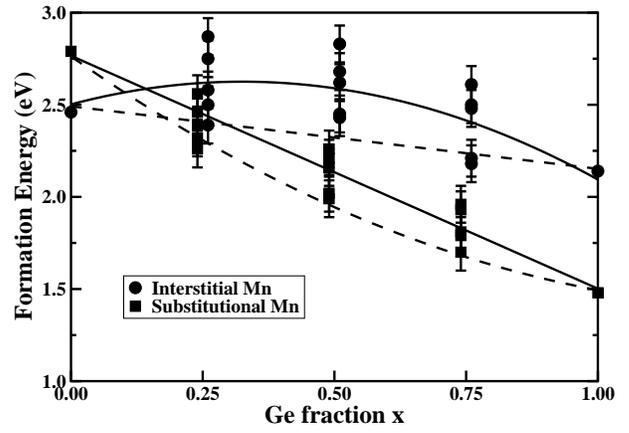}
\caption{Formation energies for interstitial and substitutional Mn in
Si$_{(1-x)}$Ge$_x$, for $x=0.25, 0.5,$ and 0.75, and for the distinct neighborhoods
Si$^{\nu}$Ge$^{4-\nu}$. The results are slightly displaced in $x$ ($x=x\pm0.01$) for clarity.
Error bars ($\pm 0.1$ eV) are estimates of the influence of vicinities beyond the nearest-neighborhood. Data
for pure Si and Ge are also shown. The continuous lines are fits to average
formations energies, whereas the long-dashed curves are fits to lowest formation energies.}
\label{fig1}
\end{figure}

All our results are presented in Table \ref{tab2} and Fig. \ref{fig1}.
For the substitutional Mn calculations, we always
removed a Si atom to place the Mn impurity, except for $x=0.5$ where a Ge atom was also removed.
These latter results are shown in parethesis in Table \ref{tab2}. As can be seen, the
formation energies are always very similar, indicating that our conclusions do not depend on Mn
replacing either a Si or a Ge atom. For interstitial Mn, the differences between the formation
energies for the local configurations Si$^{0}$Ge$^{4}$ and Si$^{4}$Ge$^{0}$ are always 0.4 eV, for all
$x$, with the lowest values occuring in Si-rich vicinities (Si$^{4}$Ge$^{0}$). For substitutional Mn this formation-energy
spread is also independent of $x$, and has a value of 0.3 eV. Ge-rich neighborhoods, however, have lower formation energies.
An important result is the fact that the lowest formation
energy for a substitutional Mn {\it is always smaller} than the lowest formation energy for an interstitial Mn.
The difference between these two lowest formation energies is 0.1 eV for $x=0.25$, 0.4 eV for $x=0.5$, and 0.5 eV for $x=0.75$.
This indicates that already for the Si$_{0.75}$Ge$_{0.25}$ alloy, there are local configurations that
make substitutional Mn the lowest energy structure. Another significant result is the fact that for $x\ge0.5$,
{\it all} substitutional configurations have formation energies that are smaller than the smallest formation
energy for interstitial Mn. The overall picture of our results can be appreciated in Fig. \ref{fig1}. Curve fittings
to both the average values (continuous curves) of the formation energies as well as to the lowest values (long-dashed
curves) of the formation energies are also presented in Fig. \ref{fig1}.
The average values curves cross at $x\simeq0.14$, whereas the lowest values curves cross at $x\simeq0.18$.
These results indicate that for $x \gtrsim 0.16\pm0.02$ there should be Ge-rich neighborhoods
in Si$_{(1-x)}$Ge$_x$ where substitutional Mn atoms becomes more stable than interstitial Mn.

At thermodynamic equilibrium, and assuming that the defects are independent of each other,
the population of Mn interstitial defects in Si$_{1-x}$Ge$_{x}$ is given by \cite{liga2}
\begin{equation}
N^I(\nu) = N_S P^\nu(x) exp(-G_f^I(\nu)/k_BT),
\end{equation}
\noindent where $N_S$ is the total number of sites in the lattice, $P^\nu(x)$ is the probability \cite{prob} for the $\nu$
configuration to exist in the $x$-concentration alloy, $G_f^I(\nu)$ is the
Gibbs free formation energy of interstitial Mn at configuration $\nu$, and $T$
is the temperature. Note that $N_S P^\nu(x)$ is the effective number of interstitial
sites of type $\nu$. Under the same assumptions, the population of Mn substitutional defects
in Si$_{1-x}$Ge$_{x}$ are given by
\begin{equation}
N^S(\nu) = N_S P^\nu(x) f(X,x) exp(-G_f^S(X,\nu)/k_BT),
\end{equation}
\noindent where $f(X,x)=(1-x)$ if $X$=Si and $f(X,x)=x$ if $X$=Ge, and $G_f^S(X,\nu)$ is the
Gibbs free formation energy of substitutional Mn at configuration $\nu$, with Mn replacing an atom
$X$ ($X$=Si,Ge). Assuming that, for a given temperature and composition, the vibrational
entropies of all defects are similar and independent on their vicinities, we can calculate
the relative population of the Mn defects as
\begin{equation}
\label{relat}
n^D(\nu,x,T) = \frac{ P^\nu(x) exp[-E_f^D(\nu)/k_BT] } {\sum_{\nu,D} P^\nu(x) exp [-E_f^D(\nu)/k_BT] }.
\end{equation}
\noindent where $D=I,S$ represents both the interstitial and substitutional defects.
For substitutional defects, since the formation energies for Mn replacing either a Ge or a Si atom are quite similar, and
given that $\sum_{X} f(X,x) = 1$, the above expression for $n^S(\nu,x,T)$ is obtained after
a summation over $X$ is performed.
\begin{table}
\caption{Relative populations for the interstitial and substitutional Mn in
Si$_{(1-x)}$Ge$_x$, for $x=0.25, 0.5$ and 0.75, and for the distinct neighborhoods
Si$^{\nu}$Ge$^{4-\nu}$. The results were obtained according to Eq. \ref{relat}, using the data from Table \ref{tab2}.}
\label{tab3}
\begin{tabular}{c|cc|cc|cc|cc|cc|cc}
\hline
\hline
       & \multicolumn{6}{c|}{T=300K}  & \multicolumn{6}{c}{T=600K}\\
      & \multicolumn{2}{c}{x=0.25} & \multicolumn{2}{c}{x=0.5} & \multicolumn{2}{c|}{x=0.75} &
 \multicolumn{2}{c}{x=0.25} & \multicolumn{2}{c}{x=0.5} & \multicolumn{2}{c}{x=0.75} \\
 $\nu$  &  \multicolumn{1}{c}{$n^I$} & \multicolumn{1}{c}{$n^S$} & \multicolumn{1}{c}{$n^I$} &
\multicolumn{1}{c}{$n^S$} & \multicolumn{1}{c}{$n^I$} & \multicolumn{1}{c|}{$n^S$} 
 &  \multicolumn{1}{c}{$n^I$} & \multicolumn{1}{c}{$n^S$} & \multicolumn{1}{c}{$n^I$} &
\multicolumn{1}{c}{$n^S$} & \multicolumn{1}{c}{$n^I$} & \multicolumn{1}{c}{$n^S$} \\
\hline
4   & 0.       & 0.     & 0. & 0.     & 0. & 0.      &  0.04  & 0.01 & 0. & 0.00 & 0. & 0.        \\
3   & 0.22   & 0.01 & 0. & 0.     & 0. & 0.      &  0.40  & 0.11 & 0. & 0.02 & 0. & 0.        \\
2   & 0.       & 0.11 & 0. & 0.00 & 0. & 0.01  &  0.      & 0.21 & 0. & 0.06 & 0. & 0.06  \\
1   & 0.       & 0.36 & 0. & 0.56 & 0. & 0.04  &  0.      & 0.18 & 0. & 0.64 & 0. & 0.18 \\
0   & 0.       & 0.30 & 0. & 0.44 & 0. & 0.95  &  0.      & 0.05 & 0. & 0.28 & 0. & 0.76 \\
\hline
\hline
\end{tabular}
\end{table}

In Table \ref{tab3} we present the relative populations for the interstitial and substitutional Mn impurities in
the Si$_{(1-x)}$Ge$_x$ alloy, for $x=0.25, 0.5$ and 0.75, for two temperatures, T=300 K and T=600 K.
In all cases the overall population of substitutional Mn is larger than the interstitial one.
At 300 K, even for $x=0.25$ we already have approximately 78~\% of Mn at substitutional sites. As the temperature
is increased to 600 K, this percentage decreases to 56~\%. The significance of this result can be appreciated
by noting that in bulk Si, at T=300 K, basically 100~\% of Mn are at interstitial sites, whereas
at T=600 K there are approximately 99.7~\% of interstitial Mn. This means that by alloying with Ge one can
revert the overall population of defects from interstitial to substitutional Mn.

In conclusion, we have shown that volume effects cannot be the main reason Mn prefers to be a subsitutional
impurity in pure Ge. Chemical effects, therefore, must clearly play an important role. Through a systematic study
of interstitial and substitutional Mn atoms in realistic
models of the Si$_{(1-x)}$Ge$_x$ alloy, we have shown that for $x \gtrsim 0.16$ the substitutional Mn in Ge-rich neighborhoods
becomes more stable than the interstitial Mn. By playing with the temperature and the concentration
$x$, and maybe also using non-equilibrium growth conditions \cite{nosso}, it should be possible
to obtain Si$_{(1-x)}$Ge$_x$ alloys with a large enough concentration of substitutional Mn atoms.
Considering that in all calculations in the alloy we obtained a Mn local moment with the same value as in the pure
crystals, and given that recent studies \cite{freeman} have shown that both Mn$_x$Si$_{1-x}$ and Mn$_x$Ge$_{1-x}$
should have similar magnetic properties, all these facts indicate that Mn:Si$_{(1-x)}$Ge$_x$ is potentially
a magnetic material like Mn$_x$Ge$_{1-x}$ \cite{mnge}, opening up in this way the road towards Si-based DMS.

This research was supported by the agencies FAPESP and CNPq. We thank CENAPAD-SP for computer time. We also
acknowledge G. M. Dalpian for useful discussions.

\end{document}